\documentclass[12pt,preprint]{aastex}

 \usepackage{epstopdf}
 \usepackage{longtable}

 \slugcomment{Accepted}

\newcommand\pv{\mbox{$p_{V}$}}
\newcommand\pIR{\mbox{$p_{IR}$}}
\newcommand\irfactor{\mbox{$p_{IR}/p_{V}$}}

\begin{document}

 \DeclareGraphicsExtensions{.pdf,.gif,.jpg}

 \title{The Population of Tiny Near-Earth Objects Observed by NEOWISE}
\author{A. Mainzer\altaffilmark{1}, J. Bauer\altaffilmark{1,2},  T. Grav\altaffilmark{3}, J. Masiero\altaffilmark{1}, R. M. Cutri\altaffilmark{2}, E. Wright\altaffilmark{4}, C. R. Nugent\altaffilmark{1}, R. Stevenson\altaffilmark{1}, E. Clyne\altaffilmark{1}, G. Cukrov\altaffilmark{1}, F. Masci\altaffilmark{2} }

 \altaffiltext{1}{Jet Propulsion Laboratory, California Institute of Technology, Pasadena, CA 91109 USA}
 \altaffiltext{2}{Infrared Processing and Analysis Center, California Institute of Technology, Pasadena, CA 91125, USA}
\altaffiltext{3}{Planetary Science Institute, Tucson, AZ USA}
\altaffiltext{4}{Department of Physics and Astronomy, UCLA, PO Box 91547, Los Angeles, CA 90095-1547 USA}

 \email{amainzer@jpl.nasa.gov}

 \begin{abstract}
Only a very small fraction of the asteroid population at size scales comparable to the object that exploded over Chelyabinsk, Russia has been discovered to date, and physical properties are poorly characterized.  We present previously unreported detections of 106 close approaching near-Earth objects (NEOs) by the Wide-field Infrared Survey Explorer mission's NEOWISE project.  These infrared observations constrain physical properties such as diameter and albedo for these objects, many of which are found to be smaller than 100 m.  Because these objects are intrinsically faint, they were detected by WISE during very close approaches to the Earth, often at large apparent on-sky velocities.  We observe a trend of increasing albedo with decreasing size, but as this sample of NEOs was discovered by visible light surveys, it is likely that selection biases against finding small, dark NEOs influence this finding.  

 \end{abstract}

 \section{Introduction}
As the products of collisional processes, small near-Earth objects (NEOs), defined as minor planets with perihelia less than 1.3 AU, are far more numerous than larger ones.  Discovering, tracking, and characterizing these objects allows us to better understand the impact hazard they pose to Earth, as well as their origins and subsequent evolution.  Because their small sizes usually make them undetectable until they are very nearby the Earth, it is often difficult for the current suite of asteroid surveys and follow-up telescopes to track them for very long.  Consequently, the fraction of the total population at small sizes that has been discovered to date remains very low.  While about 90\% of NEOs with effective spherical diameters larger than 1 km have now been discovered, the integral survey completeness drops to $\sim$25\% at 100 m, and it is likely to be $<$1\% at sizes comparable to the 17-20 m diameter NEO that exploded over Chelyabinsk, Russia on February 15, 2013 \citep{Mainzer.2011a,Harris.2008a}.  

Approximately 10,000 NEOs have been discovered to date, ~$\sim$900 of which are 1 km or larger.  The current catalog includes $\sim$3500 objects with absolute magnitude $H>22.75$ mag.  This corresponds to a diameter of 100 m or smaller assuming a geometric visible albedo \pv\ of 14\%, the average albedo for the infrared-selected NEO sample from \citet{Mainzer.2011a}, using the relationship \begin{equation}D = \left[\frac{1329\cdot 10^{-0.2H}}{p_{v}^{1/2}}\right],\end{equation} where $D$ is the effective spherical diameter \citep{Fowler.1992a, Bowell.1989a}.  For NEOs with diameters less than 20 m, corresponding to $H>$26.25 mag for \pv=14\%, there are $\sim$720 objects that have been discovered to date.  However, the true number is quite uncertain, since NEO albedos are known to range from $\sim$1\% to $>$50\% \citep{Mainzer.2011a, Trilling.2010a, Stuart.2004a}.  For NEOs 10 m and smaller, it is reasonable to assume that somewhere between 50 - 150 have been discovered.  The true numbers discovered at different size ranges depend on the objects' albedos, of course, and it is not possible to extrapolate the sample of discovered objects to the entire population without careful accounting for survey biases due to instrument sensitivity, survey cadence, weather, seeing, availability of follow-up assets, etc. \citep[c.f.][]{Jedicke.1998a, Spahr.1998a, Bottke.2002a, Mainzer.2011a, Mainzer.2012a, Grav.2011b, Grav.2012a}.  At these very small sizes, the survey completeness is very low.

Given recent interest in NEOs down to even very small sizes, it is useful to compute the average albedo for tiny NEOs discovered by visible light surveys and see how it compares to the average albedo determined for NEOs larger than 100 m by the sample returned by the \emph{Wide-field Infrared Survey Explorer's} NEOWISE project \citep{Wright.2010a, Mainzer.2011a, Mainzer.2011b}.  Because NEOWISE detected and discovered NEOs using thermal infrared wavelengths, and because new discoveries received timely ground-based follow-up, its sample was shown to be essentially albedo-insensitive \citep{Mainzer.2011a}.  The average albedo found by that sample was 14\% for NEOs 100 m and larger.  Few objects smaller than 100 m were detected by the NEOWISE automated minor planet detection software, known as the WISE Moving Object Processing System (WMOPS), which required five or more detections of objects moving at apparent on-sky velocities slower than $\sim$3.2$^{\circ}$/day.  In general, NEOs smaller than $\sim$100 m are only detected when they are quite close, resulting in significantly higher angular velocities; the smallest objects, with sizes $<$10 m, were detected when they were only 2-3 lunar distances away from Earth.  Such small objects often passed through the WISE field of view fewer than the five times required for WMOPS to detect them.  Alternately, they may have passed through the field of view more than five times, but were too faint to have been detected at least five times because of their small size or because they were trailed in the individual exposures.

A total of 429 NEOs detected by WMOPS were reported from the fully cryogenic portion of the survey, and 116 NEOs were detected and/or discovered following the partial and complete depletion of the spacecraft's cryogen \citep{Mainzer.2011a, Mainzer.2012b}.  All but a handful were larger than 100 m.  Here, we report the detection by NEOWISE of an additional 106 NEOs that were discovered by ground-based visible light surveys and made very close approaches to the Earth while WISE was observing.  These objects tend to be small and fast-moving.  This sample represents a pilot study for a future effort to conduct a wholesale search of the NEOWISE databases and images for the entire set of known minor planets; this effort will be carried out by the NEOWISE project in the near future.     

\section{Methods}
A combination of methods were used to identify known objects that were not previously identified by WMOPS in the single-exposure images.  Images were searched using the Infrared Science Archive (IRSA)/WISE Image Service\footnote{https://irsa.ipac.caltech.edu/applications/wise/} as well as by searching the WISE single-exposure source database using the Known Solar System Object Possible Association List \citep[KSSOPAL;][]{Cutri.2012a}.  Both of these tools compute an object's ephemeris and predict where it would have intersected with a WISE observation.  While WMOPS actively rejects stationary objects such as stars and galaxies before linking transient detections, neither the Image Service nor KSSOPAL discriminates between stationary objects and the asteroid, so the probability of a chance association is greater.  We rejected stationary objects by examining the WISE Atlas Images and Catalogs, which combine all possible exposures at a given location to produce a deeper image, following the methods described in \citet{Mainzer.2011a, Mainzer.2011c}.    

Of the currently known NEOs with $H>26.25$ mag, the median observational arc length spans only $\sim$3 days, and only a few dozen exceed 20 days.  The vast majority of these objects are therefore lost, frustrating efforts to locate them in the NEOWISE data.  However, many of the objects that were discovered by ground-based surveys during the WISE survey phase did pass through the WISE field of view near 90$^{\circ}$ solar elongation and were bright enough to be detected, albeit often with fewer than five observations.  Only those objects whose astrometric uncertainties were very small (less than a few arc seconds) at the time of their passage through the WISE field of view were included in this analysis, with the exception of a handful of extremely close-approaching NEOs with larger uncertainties that were identifiable by color and morphology (the images were slightly trailed).  Astrometric uncertainties were taken from the MPC's ephemeris service.  For objects that were stacked, only objects with very low astrometric uncertainty (less than $\sim$3 arcsec) were used.  Future work will extend to identifying and stacking objects with larger astrometric uncertainties.

The WISE instrument used three beamsplitters to collect images in all four bands simultaneously (3.4, 4.6, 12 and 22 $\mu$m, hereafter W1, W2, W3, and W4).  The exposure time in all four WISE bands was set to 8.8 sec in bands W3 and W4 and 7.7 sec in bands W1 and W2.  An NEO moving faster than $\sim 17^{\circ}$/day could thus be trailed in the 6.5 arcsec W3 beam.  All candidate images were visually inspected for evidence of trailing.  If the images were trailed, aperture magnitudes reported by the data reduction pipeline (denoted w1mag\_x, w2mag\_x, w3mag\_x, and w4mag\_x, where 1-4 represents the wavelength band, 1-4, and x represents the number of the aperture that encircled the source, 1-8) were used instead of photometry derived using point-spread function fits; see Section IV.4.c of the WISE Explanatory Supplement \citep{Cutri.2012a} for a detailed description of the WISE pipeline photometry.  Aperture curves-of-growth were constructed for the trailed objects, and the aperture at which the curve converged was selected.   

\begin{figure}
\figurenum{1}
\includegraphics[width=6in]{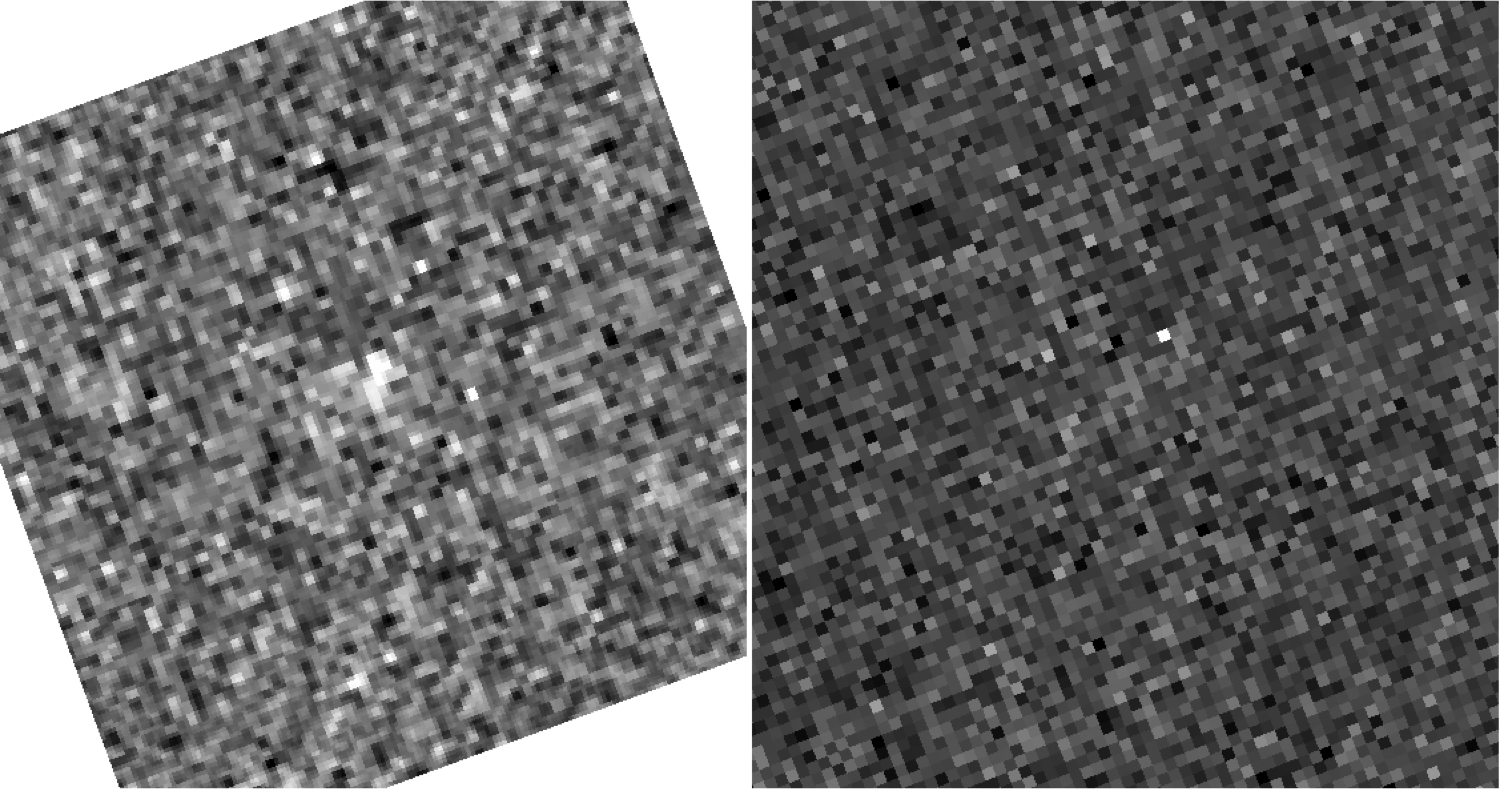}
\caption{\label{fig:stack} Left: The NEO 2010 FK (center of image) was only detected in W3 by creating a moving coadd that combined 13 exposures, as it fell just below the detection threshold in the single-frame images (right, center of image).   }
\end{figure}

For objects that were too faint to be detected reliably in single exposures, the single exposure images were coadded in the moving frame of the asteroid, aligning the frames using each object's ephemeris, with the ICORE (Image Co-addition with Optional Resolution Enhancement) stacking algorithm \citep{Masci.2009a, Masci.2013a}.  The ICORE algorithm includes outlier rejection and resamples the stacked image to a pixel scale of 1 arcsec/pixel.  The algorithm requires a minimum of five images to ensure adequate pixel outlier rejection; all coadded objects in this analysis exceeded this number of images.  An example of an object recovered in the NEOWISE data by stacking with ICORE is shown in Figure \ref{fig:stack}, demonstrating the utility of this technique for obtaining infrared detections of minor planets that fell just below the single frame detection threshold; similar success with the extended object comet 17P/Holmes was shown in \citet{Stevenson.2013a}.  For the stacked objects, aperture photometry was performed using radii of 11, 11, 11, and 22 arcsec, respectively, in the four WISE bands.  Table 1 lists the magnitudes and apertures used for each object.

\subsection{Thermal Modeling and Error Analysis}
\begin{figure}
\figurenum{2}
\includegraphics[width=3in]{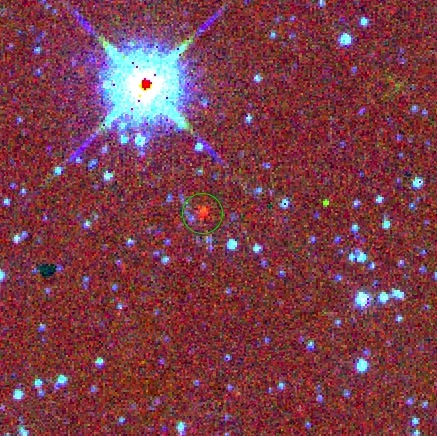}
\caption{\label{fig:K10G07H} The NEO 2010 GH$_{7}$ (circled in green) was detected only once by NEOWISE due to its high apparent on-sky velocity in bands W2, W3, and W4.  This object is on the list of accessible targets for potential human exploration; it makes close approaches to Earth every $\sim$5 years.  Blue = band W1; green = W2; red = W3.   }
\end{figure}

The Near-Earth Asteroid Thermal Model \citep[NEATM;][]{Harris.1998a} was used to determine diameters and albedos for most objects.  The color corrections and modifications that account for the observed discrepancy between red and blue calibrators noted in \citet{Wright.2010a} were applied following the methods described in \citet{Mainzer.2011c}.  Table 2 shows the thermal fit results for the 106 NEOs that were recovered here; Figure \ref{fig:K10G07H} shows an example of a small ($\sim$8 m) NEO identified in the NEOWISE data using the IRSA/WISE Image Service and KSSOPAL.  

As described in \citet{Mainzer.2011a} and \citet{Mainzer.2011c}, the so-called beaming parameter $\eta$ employed by the NEATM was fit when two or more thermally-dominated infrared bands were available.  Since an NEO's flux at 3.4 and 4.6 $\mu$m generally consists of a mix of thermal emission and reflected sunlight, it was necessary to fit for the reflectivity at these wavelengths.  Albedo was assumed to be the same for both 3.4 and 4.6 $\mu$m.  This simplifying assumption may not always be valid.  \citet{Grav.2012a} and \citet{Grav.2012b} have shown that albedo varies between 3.4 and 4.6 $\mu$m for Hilda group asteroids and Jovian Trojans, although the NEOs' smaller heliocentric distances and warmer temperatures means that the 4.6 $\mu$m band is often thermally-dominated, lessening the effect of albedo at 4.6 $\mu$m on the total flux.  The infrared flux, $p_{IR}=p_{3.4 \mu m}=p_{4.6 \mu m}$, could only be fit when the flux at 3.4 $\mu$m was dominated by reflected sunlight, which depends on both reflectivity and heliocentric distance.    

The best-fit values for diameter, visible geometric albedo \pv, infrared albedo \pIR, and beaming parameter $\eta$ were determined using a least-squares optimization that accounted for the measurement uncertainties for bands W1, W2, W3, W4, absolute magnitude $H$, and phase curve slope parameter $G$. $G$ was generally taken to be 0.15$\pm$0.1 unless a measurement was available \citep{Bowell.1989a}.  Errors on $H$ ($\sigma _{H}$) were assumed to be 0.3 mag, although in some cases, examination of the MPC observations file revealed the uncertainty in $H$ to be considerably larger; in these cases, $H$ was assumed to be unknown and was not used in the thermal fit.  Poorly known $H$ and $G$ values continue to be a persistent difficulty that sometimes inhibits precise determination of albedo, although this problem does not much affect the determination of diameter when using radiometric methods such as NEATM.  

Although there is evidence that systematic biases in $H$ magnitudes may be present in the existing asteroid databases such as MPC and \emph{Horizons} that derive $H$ using measurements made primarily for improvements to astrometry rather than accurate photometry, targeted measurements of $H$ \citep[e.g.][]{Pravec.2012a} have concentrated on objects with $H\leq$21 mag.  We therefore chose not to adopt a blanket offset to all $H$ values and instead bounded the errors by assuming $\sigma _{H}=0.3$ mag.  The errors in diameter, \pv, \pIR, and $\eta$ were determined through the use of Monte Carlo trials that varied the measured values of W1, W2, W3, W4, $H$, and $G$ by their respective errors.  If \irfactor\ could not be fit, the default value was taken to be the average value determined in \citet{Mainzer.2011a} for NEOs, or 1.6$\pm$1.0.  

Another source of error is that the slope parameter $G$ is known to vary with taxonomic type; C-complex asteroids typically have lower $G$ values, closer to 0.05-0.10, whereas S-complex asteroids have $G$ closer to 0.2, and the highest albedo classes such as E-types have $G\sim$0.4 \citep{Harris.1989a, Harris.1989b, Lagerkvist.1990a, Oszkiewicz.2012a}.  Our generic assumption of $G=0.15$ can therefore yield $H$ values that are too high for C-type objects, which in turn result in overly low values of \pv.  Similarly, our assumed $G$ value of 0.15 is likely to be too low for S-complex objects, particularly E- and V-types, resulting in $H$ values that are erroneously low.  These effects will be more pronounced at the high phase angles at which NEOWISE typically observed small NEOs and Hungarias.  Erroneous values of $G$ and $H$ do not much affect diameters derived from thermal measurements, but rather albedo.  For example, assuming that 2010 TN$_{4}$ has $G=0.05$ instead of $G=0.15$ results in an offset to $H$ of -0.28 mag at its NEOWISE-observed phase angle of 80.6$^{\circ}$.  The derived diameter (18 m) is unchanged, but \pv=0.071 instead of 0.054.  

However, without knowledge of an object's spectral type, we cannot know for certain which value of $G$ to choose if it has not been directly measured.  In our previous works \citep[e.g.][]{Mainzer.2011d, Mainzer.2012c}, we found that there is not a perfect correlation between albedo and taxonomic type.  We therefore chose not to apply offsets to $G$ based on assumed spectral type (which in most cases we could only assume based on albedo - a circular argument).  Instead, we model and bound the errors caused by imperfect knowledge of $G$ by assuming that $G$ varies by $\pm$0.10 in the Monte Carlo trial fits for each object.  It is important to note that errors in $G$ and $H$ could cause systematic offsets in albedo, with lower albedo objects' albedos sometimes being too low, and higher albedo objects' albedos coming out too high.  For this reason, direct measurements of $H$, $G$, and spectral type for individual objects would be useful, particularly when determining albedos of objects observed at high phase angles.  Nevertheless, a major benefit of radiometric fits to infrared observations is that while errors in $H$ and $G$ can affect albedo, they cause little change to diameter.

\subsection{Results}

\begin{figure}
\figurenum{3}
\includegraphics[width=6in]{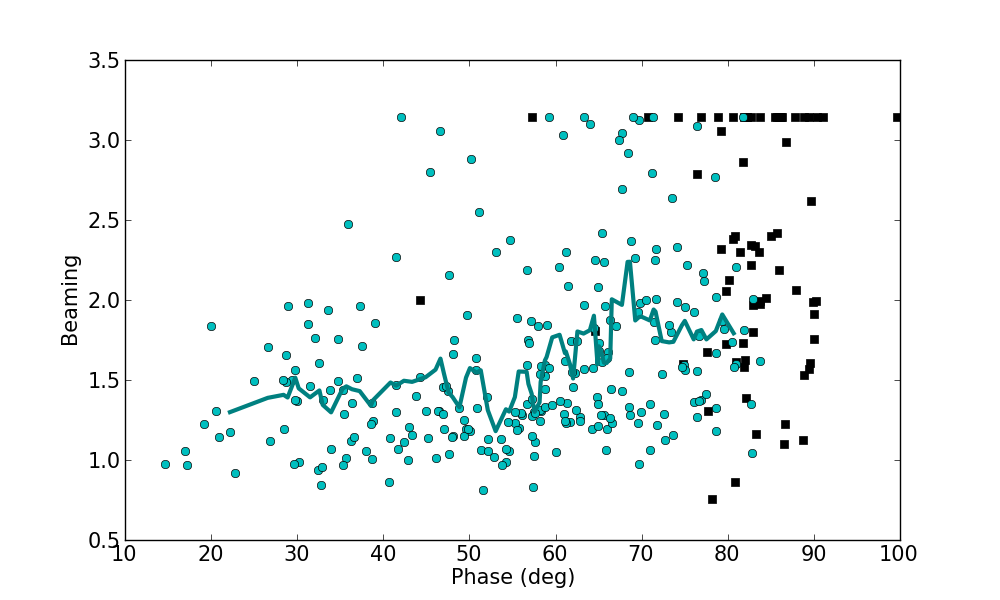}
\caption{\label{fig:beaming} Beaming parameter vs. phase angle for objects observed in two or more thermal wavelengths.  The small, close-approaching NEOs that were detected by NEOWISE using the KSSOPAL and WISE Image Service tools (black squares) were usually observed at very high phase angles; the beaming parameter $\eta$ required was significantly larger than the average value for NEOs observed at lower phase angles (cyan points; cyan line shows running median).  Cyan points taken from \citet{Mainzer.2011a}.  Only objects with more than one thermally-dominated band available are plotted as these are the objects for which $\eta$ can be fit; the maximum value of $\eta=\pi$.}
\end{figure}

Figure \ref{fig:beaming} shows the beaming distribution for the small, close-approaching NEOs recovered from the NEOWISE data compared with the sample of 429 NEOs detected by WMOPS during the fully cryogenic portion of the mission.  The close-approaching NEOs tend to have been observed near 90$^{\circ}$ phase, because the WISE spacecraft only observed near 90$^{\circ}$ solar elongation \citep{Wright.2010a}.  Thus, the objects were typically observed near their local terminators.  The NEATM assumes zero flux contribution from the night side of the asteroids.  Because small NEOs are frequently rapidly rotating with rotation periods much less than the cooling timescale \citep{Pravec.2008a, Hergenrother.2011a, Statler.2013a}, the approximation of zero night side flux is likely inappropriate in many cases.  Indeed, the beaming parameter $\eta$, which is used by the NEATM to compensate for the ``beaming" effect of radiation observed near zero phase angle, converges to its theoretical maximum value of $\pi$ in many cases when more than one thermally dominated band is available.  The average value found for the small NEOs reported in this work was 2.0$\pm$0.5, higher than the 1.4$\pm$0.5 reported for all NEOs over a wide range of phase angles in \citet{Mainzer.2011b}.  Therefore, if $\eta$ could not be fit, a default value of 2.0$\pm$0.5 was used for objects observed close to 90$^{\circ}$ phase angle.  However, we note that while a fit yielding the theoretical maximum value of $\eta=\pi$ may indeed indicate high thermal inertia and/or rapid rotation, it could also be that the NEATM's assumption of zero nightside flux is inappropriate for asteroids observed at very high phase angles.  At high phase angles, a substantial portion of the nightside hemisphere is visible, so the approximation of zero nightside flux contribution may be poor.  Furthermore, for the very smallest objects, the heat conduction length scale begins to approach the object size.    

In cases where $\eta = \pi$ resulted from the NEATM fit, the Fast Rotating Model \citep[FRM;][]{Lebofsky.1978a, Lebofsky.1989a, Veeder.1978a} was used instead.  In the FRM, the asteroid is assumed to be rotating rapidly compared to its cooling timescale, so that the temperature is isotropic with respect to longitude and varies only with local latitude, assuming that the object's rotational axis is perpendicular to the Earth-Sun line.  The FRM uses fast rotation to smooth out longitudinal temperature variations, although it still has latitudinal temperature variations. The timescale for these to change is seasonal, so thermal conduction (especially in a solid boulder) could lead to uniform temperature in longitude.  When using the FRM, we made the assumption that the rotational pole is perpendicular to the plane defined by the Sun, Earth, and object.  This assumption may be reasonable given that there is evidence that the non-gravitational thermal pressure torques exerted by the YORP effect are thought to drive small asteroids to this state.  For objects in the 10 - 100 m size range, the mean time between spin axis reorientation due to collisions is thought to be much longer than the timescale over which YORP will reshape spin states \citep{Farinella.1998a, Rossi.2009a}.  Using the FRM instead of the NEATM usually has the effect of shrinking an object's best-fit effective spherical diameter, since the flux that was distributed across only one hemisphere using the NEATM is now spread across the entire visible area; diameter must therefore shrink to conserve the emitted energy.   

With so few observations (in some cases only one), the rotational lightcurves of many NEOs in this sample were not well-sampled.  For the WMOPS-detected sample, the WISE observational cadence typically resulted in 10-12 observations of each object collected over a $\sim$36 hour span.  The question is the extent to which diameters determined using only sparse detections are representative of the actual effective spherical diameter.  As shown in figure 5 in \citet{Mainzer.2011a}, most NEOs detected by NEOWISE had lightcurve amplitudes of $\sim$0.4 mag or less.  

Two effects come into play when one attempts to use sparse detections near the sensitivity limit to investigate a population.  First, there is a systematic bias in flux measurements that are made very close to the noise limit of a detector.  Random noise can scatter flux above the detection threshold, making a source appear brighter.  This manifestation of the Eddington bias was observed with WISE's predecessor, the \emph{Infrared Astronomical Satellite}\footnote{http://irsa.ipac.caltech.edu/IRASdocs/exp.sup/ch11/J.html}.  Second, asteroids rotate, and they can be elongated.  The tendency would be to detect a rotating, elongated body when it is closer to presenting its maximum surface area to the observer.  Both of these effects combine to produce an overestimate of fluxes for small bodies observed by WISE only a handful of times.    

\begin{figure}
\figurenum{4}
\includegraphics[width=6in]{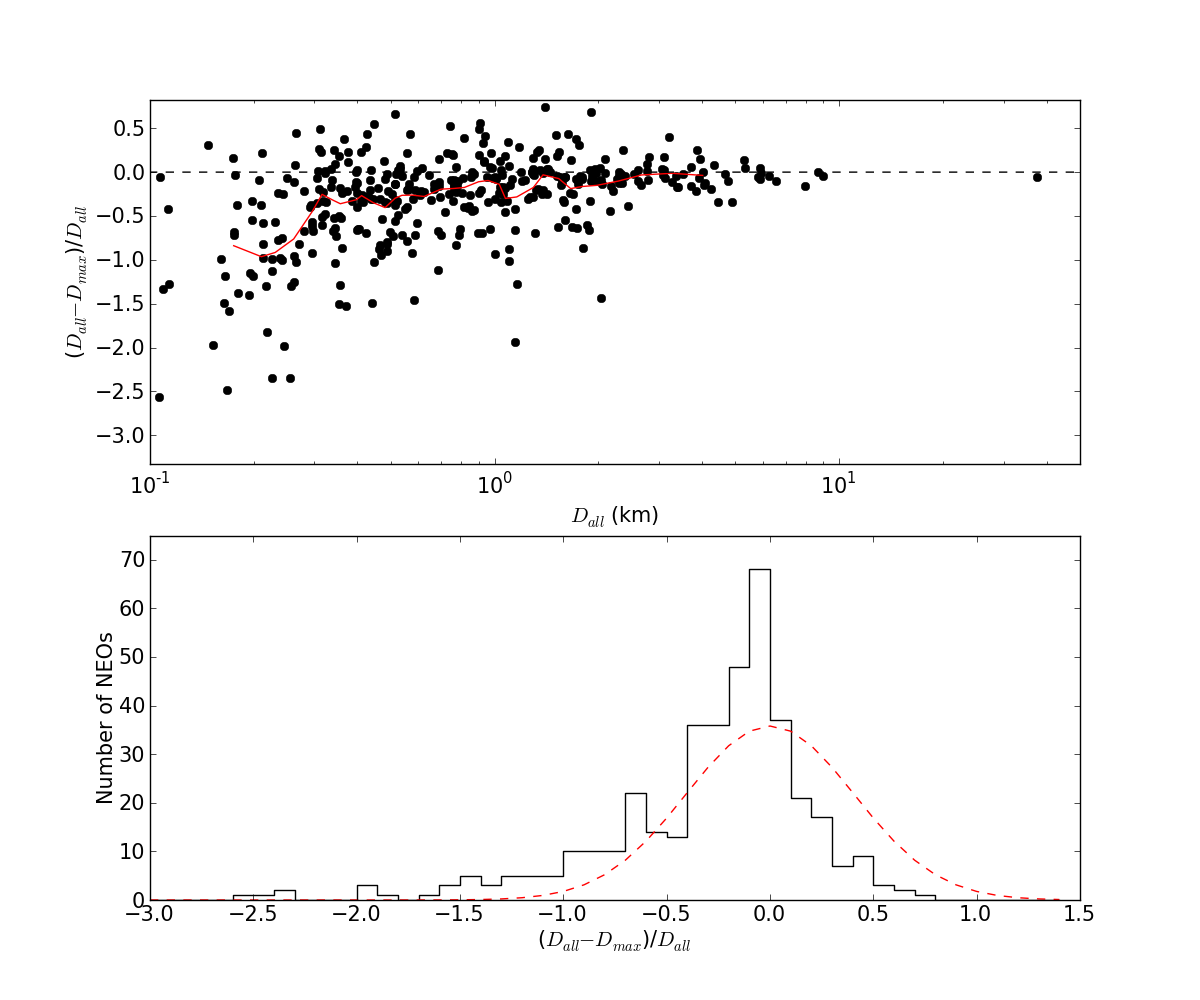}
\caption{\label{fig:diam} Top: Comparison between NEATM diameter fits derived using the full set of NEOWISE measurements for each NEO vs. fits derived using only the brightest measurement in W3 for each object; the red line shows a running median.  Bottom: Histogram of diameter differences between full lightcurve fits and maximum brightness fits; a Gaussian is overplotted (red dashed line).}
\end{figure}

In an effort to better understand the effects of sparsely sampled infrared lightcurves on thermal model outputs, a comparison was made between fits using the entire set of NEOWISE thermal measurements for each object and fits derived using only the single brightest point per object.  Because WISE observes in all four bands simultaneously using beamsplitters, we selected the time at the maximum of the lightcurve in the band with the highest signal-to-noise ratio, then used all bands available at that time.  Figure \ref{fig:diam} (top) shows $\Delta D = (D_{all} - D_{max})/D_{all}$, where $D_{all}$ is the diameter derived using all points, and $D_{max}$ are the diameters resulting from the maximum brightnesses, as a function of $D_{all}$.  Over most size ranges, the dispersion between $D_{max}$ and $D_{all}$ is typically a factor of $\sim$1.3, although it worsens at smaller sizes.  In most cases, the tendency is unsurprisingly to overestimate the diameter when using only the observations at the maximum of a lightcurve.  However, at smaller sizes in particular, $D_{max}$ can be observed in Figure \ref{fig:diam} to be approximately double $D_{all}$.   

Unfortunately, this analysis cannot account for the possibility of real variations in the shapes of NEOs at sizes smaller than the WMOPS-selected sample shown in Figure \ref{fig:diam}.  Very small NEOs might have shapes that are either more round or more irregular than larger objects, depending on their origins and subsequent evolution.  The observed amplitude of an asteroid's light curve depends on the interplay between the shape of the object, the orientation of the rotation pole with respect to the line of sight, and the phase angle of the object at the time of observation when using reflected light.  To first order, the larger the axial ratios of the object, the larger the amplitude of the light curve will be.  However, if the sub-observer point is near the rotation pole, the light curve amplitude will be reduced.

Amplitudes larger than two magnitudes have been seen at optical wavelengths for small NEOs that are a result of very elongated shapes \citep[e.g. 2009 UU$_{1}$;][]{Warner.2009a}.  However, small amplitudes are frequently observed by WISE, implying either a shape close to spherical or an observing geometry looking along the rotation pole.  Nevertheless, the analysis described above suggests that the sizes derived from sparse observations can serve as valuable upper limits to NEO sizes, and it is therefore quite likely that many of the objects presented in Table 2 are somewhat smaller than their derived sizes.  The error bars reported in Table 2 are the formal errors resulting from the Monte Carlo trials that vary the WISE magnitudes, $H$, and $G$ by their respective error bars, and they do not include the systematic errors that may be associated with the sparseness of some of the lightcurve sampling.  A typical systematic error for the sizes given in Table 2 is likely to be of order the $\sim$30\% shown in Figure \ref{fig:diam}, although in some cases, it is possible that the derived sizes are quite different.  

The structure of asteroids smaller than 100 m is uncertain.  \citet{Pravec.2008a} have shown that while asteroids larger than a few hundred meters show a spin limit coincident with the theoretical maximum rotation rate of a gravitationally bound rubble pile, smaller asteroids typically rotate well above this limit, which was interpreted as evidence that these objects were monolithic.  \citet{Scheeres.2010a} showed that cohesive forces on small grains ($<1~$cm) can be sufficient to allow small rubble piles to rotate at rates above the spin barrier without disrupting.  However, in the process of forming these objects, any monolithic components of the original body could be ejected onto independent orbits, becoming NEOs in their own right.  Thus, there may be different kinds of small NEOs.  Monolithic fragments created by the breakup of larger bodies will have shape distributions determined by the impact and fracture processes of the collisions that generated them, which may vary among collisions.  Objects held together by cohesive forces may have a range of shapes allowed by the force balance, from nearly spherical to oblong.  While it is beyond the scope of this work to constrain these shape distributions and the division between monolithic and cohesive objects, we can regard the diameters derived from sparsely sampled lightcurves as, at worst, useful upper limits.  

\emph{2009 BD.}  One of the objects shown in Table 2, 2009 BD, passed through the WISE fields of view but was not detected even when the images were stacked.  However, it is possible to use the images to set an upper limit on its size from the non-detection.  We extracted magnitudes from 13 images coadded using ICORE, taken on June 13 and 14 of 2010 (during the fully cryogenic mission phase) at a phase angle of 88$^{\circ}$, using aperture photometry with a sample radius of 11 arcsec.  The counts yielded a $5-\sigma$ magnitude limit of 11.77 for W3, our most sensitive band for NEOs, and 7.59 mag for W4.  Given that 2009 BD was discovered by a visible light telescope (the Catalina Sky Survey), Figure \ref{fig:diam_alb} suggests that it is more likely to have a higher albedo than a lower albedo.  Using the NEATM fit code, this magnitude limit corresponds to a 5-$\sigma$ effective spherical diameter upper limit of $\sim$14.5 m, assuming \pv=0.2 and $\eta$=2.0; using \pv=0.04 produces an upper limit of 14 m.  The FRM results in a 5-$\sigma$ upper limit of 8 m.

\subsection{Discussion}
\begin{figure}
\figurenum{5}
\includegraphics[width=6in]{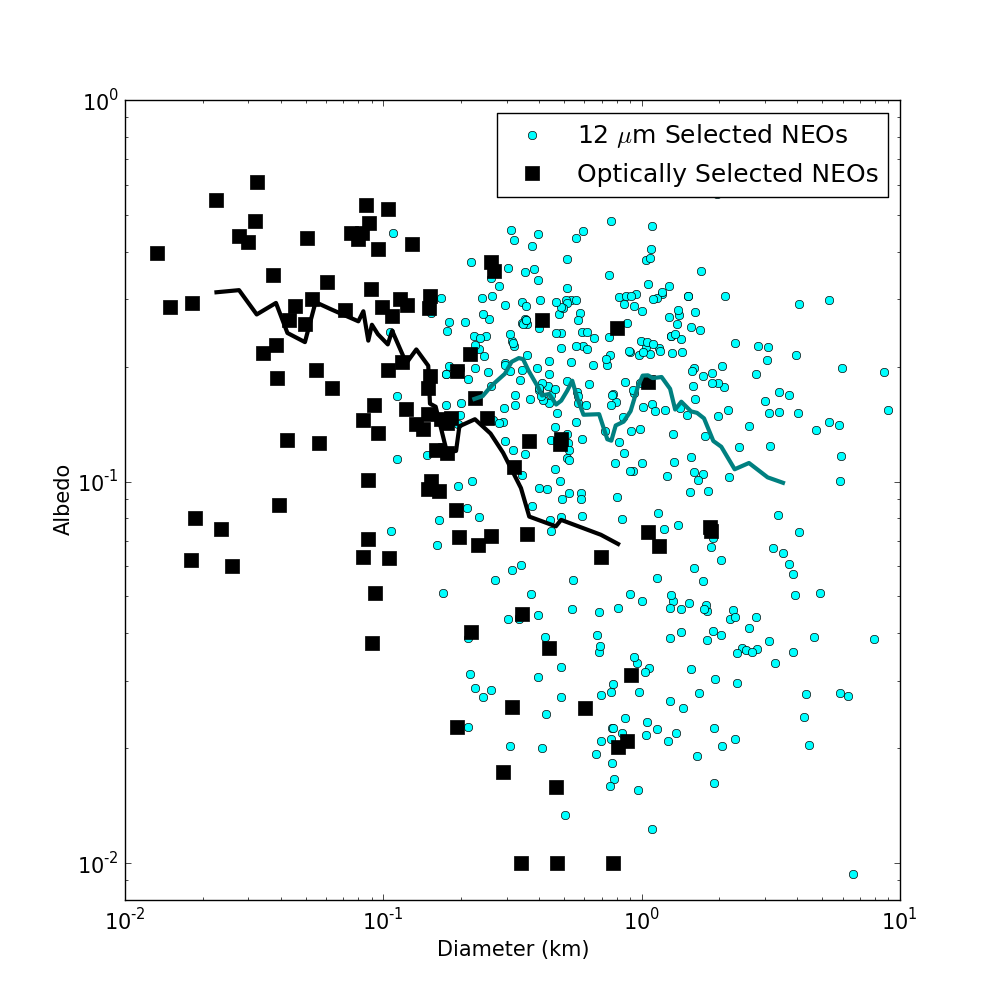}
\caption{\label{fig:diam_alb} The sample of objects recovered from the NEOWISE dataset using the KSSOPAL and WISE Image Service tools were all discovered by visible light ground-based surveys (black squares).  The albedo distribution of these objects is distinctly different from the sample of NEOs that were selected using the WMOPS algorithm working on 12 $\mu$m NEOWISE data (cyan circles).  Because the WMOPS algorithm treated new discoveries the same way as recoveries of previously known objects, and because asteroids' thermal fluxes depend only weakly on albedo, the 12 $\mu$m sample is albedo-insensitive.  The optically-selected sample's albedo distribution increases sharply with decreasing size (black line), whereas the albedo distribution of the infrared-selected sample (cyan line) remains essentially unchanged with decreasing size. }
\end{figure}

Figure \ref{fig:diam_alb} shows the diameters and albedos of the optically-selected sample of small NEOs recovered from the NEOWISE data compared with the infrared-selected sample drawn using WMOPS from \citet{Mainzer.2011a}.   All recoveries were discovered by visible light telescopes, which are preferentially biased against discovering small, dark NEOs.  The infrared-selected sample's albedo distribution is essentially flat with diameter, whereas the optically-selected sample's albedos increase with decreasing diameter.  While there may be real physical changes in albedo with diameter, the visible light survey biases against discovering small, dark NEOs complicate efforts to determine the true albedo distribution of the population of small NEOs.  It is also apparent that the current suite of visible light survey telescopes is not efficient at discovering very tiny low albedo NEOs.  

While one might be tempted to use the albedo distribution for tiny NEOs shown in Figure \ref{fig:diam_alb} to extrapolate to the total number of NEOs in the population \citep[c.f.][]{Mainzer.2011a,Mainzer.2012a}, we caution that this is not an infrared-selected sample with well-determined survey biases.  All objects presented here were discovered by ground-based visible light surveys that are subject to the effects of weather and seeing variations, in addition to the NEOWISE detection biases.  Debiasing this sample  requires careful modeling of all of these selection effects, and for objects in this size range, infrasound measurements of small meteors and cratering studies should be included \citep[c.f. the lunar impact flash studies such as][]{Oberst.2012a, Suggs.2008a} to provide a realistic assessment of their true numbers, sizes, and albedos.  

\section{Conclusions}
We have reported sizes, albedos, and thermal model parameters such as beaming for 106 NEOs, roughly half of which are smaller than 100 m in effective spherical diameter, and the smallest of which are $\sim$8 m.  These objects generally were not detected by the automated WMOPS survey because they were too faint in the individual images or were not detected the required five times for the pipeline to trigger on them.  Many of the objects were extremely close to Earth when they were observed by WISE, with $\sim$10 m NEOs being detectable only when they approached within several lunar distances.  Because close NEOs tend to move with large apparent velocities, many of the objects were only detected a handful of times; approximately one-third of the sample reported here was detected only once.  While radiometric thermal models are typically able to constrain the sizes of asteroids when lightcurves are well-sampled, the results of models using these sparsely-sampled lightcurves must be regarded with caution.  We have attempted to estimate the degree to which thermal models of NEOs using sparse data will err in their predictions of effective spherical diameter.  We find that in most cases, typical errors are $\sim$30\% if one assumes that the distribution of shapes for NEOs smaller than $\sim$100 m is similar to those larger than 100 m.  However, in some cases, the size is overestimated by factors of several.  Nonetheless, sparse infrared data can still provide useful estimates of effective spherical diameters.

We have shown that our sample of small NEOs displays a marked trend of increasing albedo with decreasing diameter, but given that the optical surveys that discovered this sample are biased against finding small, dark, faint NEOs, this is not surprising.  These data alone are insufficient to determine whether or not this result represents a real physical trend or merely selection effects, as the optical surveys are highly incomplete at these small sizes.   Selection biases must be carefully disentangled using models of survey performance.  It is, however, clear that much work remains to be done to discover and characterize the population of very small NEOs.

\section{Acknowledgments}

\acknowledgments{This publication makes use of data products from the \emph{Wide-field Infrared Survey Explorer}, which is a joint project of the University of California, Los Angeles, and the Jet Propulsion Laboratory/California Institute of Technology, funded by the National Aeronautics and Space Administration.  This publication also makes use of data products from NEOWISE, which is a project of the Jet Propulsion Laboratory/California Institute of Technology, funded by the Planetary Science Division of the National Aeronautics and Space Administration.  We thank our referee, Dr. Alan Harris of Pasadena, for his helpful comments that materially improved this manuscript.  We gratefully acknowledge the extraordinary services specific to NEOWISE contributed by the International Astronomical Union's Minor Planet Center, operated by the Harvard-Smithsonian Center for Astrophysics, and the Central Bureau for Astronomical Telegrams, operated by Harvard University.  We also thank the worldwide community of dedicated amateur and professional astronomers devoted to minor planet follow-up observations. This research has made use of the NASA/IPAC Infrared Science Archive, which is operated by the California Institute of Technology, under contract with the National Aeronautics and Space Administration. }

 \clearpage

\clearpage

\begin{deluxetable}{lllllll}
\tabletypesize{\tiny}
\tablecolumns{7}
\tablecaption{\scriptsize{Observed WISE magnitudes for each of the NEOs presented here, including the modified Julian date (MJD).  Objects that were not detected at a particular wavelength represent $2-\sigma$ upper limits \citep{Cutri.2012a}.  A value of "--" indicates that no data were available at that wavelength.  The final column gives the aperture radius in arcsec used for aperture photometry; ``0" indicates that the pipeline profile fit photometry was used.   Objects marked with * were stacked in order to recover their fluxes; for the objects, the stacked image photometry is reported.}}
\tablewidth{0pt}
\tablehead{\colhead{Name} & \colhead{MJD} &\colhead{W1 (mag)} & \colhead{W2 (mag)} & \colhead{W3 (mag)}  & \colhead{W4 (mag)} & \colhead{Aperture}}
\startdata
  85989 & 55310.732616 & $>$16.882 & $>$15.004 & 9.264 $\pm$ 0.081 & 7.098 $\pm$ 0.260 &  0\\
  85989 & 55310.86492 & $>$16.703 & 14.323 $\pm$ 0.175 & 8.765 $\pm$ 0.055 & 6.734 $\pm$ 0.198 &  0\\
  85989 & 55310.997225 & $>$16.695 & 14.525 $\pm$ 0.266 & 8.247 $\pm$ 0.035 & 6.260 $\pm$ 0.124 &  0\\
  85989 & 55311.129529 & 16.623 $\pm$ 0.431 & 14.170 $\pm$ 0.162 & 8.195 $\pm$ 0.035 & 5.864 $\pm$ 0.106 &  0\\
  85989 & 55311.261833 & 16.237 $\pm$ 0.300 & 14.507 $\pm$ 0.212 & 8.598 $\pm$ 0.052 & 6.514 $\pm$ 0.161 &  0\\
  85989 & 55311.327922 & $>$16.373 & 14.234 $\pm$ 0.171 & 8.491 $\pm$ 0.044 & 6.320 $\pm$ 0.142 &  0\\
  85989 & 55311.394137 & $>$17.032 & $>$15.588 & 8.842 $\pm$ 0.056 & 7.073 $\pm$ 0.294 &  0\\
  85989 & 55311.460226 & $>$16.506 & 13.843 $\pm$ 0.126 & 8.190 $\pm$ 0.037 & 6.145 $\pm$ 0.113 &  0\\
\enddata
\end{deluxetable}

\clearpage
\clearpage

\begin{deluxetable}{lllllll}
\tabletypesize{\tiny}
\tablecolumns{7}
\tablecaption{\scriptsize{Thermal fit results for the 106 new NEO detections reported in this work. This table contains the preliminary thermal fit results based on the first-pass version of the WISE data processing as described in the text. The columns contain object name, $H$ magnitude, phase curve slope parameter $G$, diameter, visible albedo \pv, beaming parameter $\eta$, infrared albedo $p_{IR}$, and number of observations in each of the four WISE bands.  The $1 -\sigma$ errors presented here were statistically generated using Monte Carlo modeling. WISE fluxes, absolute magnitude $H$, and $G$ were varied by their $1-\sigma$ error bars, as well as beaming ($\eta$) and $p_{IR}$ when these two parameters could not be fit.  The quoted precision for each parameter follows the object with the most significant figures for the error on that value in the  table.  The statistical errors on diameter and \pv\ for each object in the table should be added in quadrature to the  systematic errors described in the text and discussed in \citet{Mainzer.2011c}. If $\eta=\pi$, the FRM was used instead of NEATM.  Objects with multiple entries were observed at multiple epochs, which were fit separately.  }}
\tablewidth{0pt}
\tablehead{\colhead{Object} & \colhead{$H$} &\colhead{$G$} & \colhead{$D$ (km)} & \colhead{\pv}  & \colhead{$\eta$} & \colhead{$p_{IR}$}  }
\startdata
  85989 & 17.10 & 0.15 &    1.834 $\pm$    0.018 &    0.076 $\pm$    0.017 &    1.674 $\pm$    0.025  &    0.072 $\pm$    0.027\\
  85989 & 17.10 & 0.15 &    1.853 $\pm$    0.077 &    0.074 $\pm$    0.018 &    2.001 $\pm$    0.151  &    0.994 $\pm$    0.110\\
  88254 & 17.60 & 0.15 &    0.800 $\pm$    0.012 &    0.252 $\pm$    0.035 &    1.309 $\pm$    0.046  &    0.352 $\pm$    0.150\\
  D7032 & 16.60 & 0.15 &    1.055 $\pm$    0.322 &    0.183 $\pm$    0.325 &    2.000 $\pm$ 0.500   &    0.310 $\pm$    0.288\\
  F2742 & 19.10 & 0.15 &    0.413 $\pm$    0.005 &    0.264 $\pm$    0.011 &    1.400 $\pm$ 0.500   &    0.276 $\pm$    0.086\\
  F4590 & 21.70 & 0.15 &    0.086 $\pm$    0.002 &    0.530 $\pm$    0.076 &    1.400 $\pm$ 0.500   &    0.848 $\pm$    0.114\\
  M6554 & 19.50 & 0.15 &    0.482 $\pm$    0.007 &    0.125 $\pm$    0.024 &    3.055 $\pm$    0.060  &    0.200 $\pm$    0.038\\
  N0549 & 20.60 & 0.15 &    0.217 $\pm$    0.004 &    0.216 $\pm$    0.046 &    2.402 $\pm$    0.072  &    0.346 $\pm$    0.073\\
\enddata
\end{deluxetable}

\clearpage


\clearpage
\begin{thebibliography}{41}
\expandafter\ifx\csname natexlab\endcsname\relax\def\natexlab#1{#1}\fi

\bibitem[{{Bottke} {et~al.}(2002){Bottke}, {Morbidelli}, {Jedicke}, {Petit},
  {Levison}, {Michel}, \& {Metcalfe}}]{Bottke.2002a}
{Bottke}, W.~F., {Morbidelli}, A., {Jedicke}, R., {Petit}, J.-M., {Levison},
  H.~F., {Michel}, P., \& {Metcalfe}, T.~S. 2002, \icarus, 156, 399

\bibitem[{{Bowell} {et~al.}(1989){Bowell}, {Hapke}, {Domingue}, {Lumme},
  {Peltoniemi}, \& {Harris}}]{Bowell.1989a}
{Bowell}, E., {Hapke}, B., {Domingue}, D., {Lumme}, K., {Peltoniemi}, J., \&
  {Harris}, A.~W. 1989, in Asteroids II, ed. {R.~P.~Binzel, T.~Gehrels, \&
  M.~S.~Matthews}, 524--556

\bibitem[{{Cutri} {et~al.}(2012){Cutri}, {Wright}, {Conrow}, {Bauer},
  {Benford}, {Brandenburg}, {Dailey}, {Eisenhardt}, {Evans}, {Fajardo-Acosta},
  {Fowler}, {Gelino}, {Grillmair}, {Harbut}, {Hoffman}, {Jarrett},
  {Kirkpatrick}, {Leisawitz}, {Liu}, {Mainzer}, {Marsh}, {Masci}, {McCallon},
  {Padgett}, {Ressler}, {Royer}, {Skrutskie}, {Stanford}, {Wyatt}, {Tholen},
  {Tsai}, {Wachter}, {Wheelock}, {Yan}, {Alles}, {Beck}, {Grav}, {Masiero},
  {McCollum}, {McGehee}, {Papin}, \& {Wittman}}]{Cutri.2012a}
{Cutri}, R.~M., {Wright}, E.~L., {Conrow}, T., {Bauer}, J., {Benford}, D.,
  {Brandenburg}, H., {Dailey}, J., {Eisenhardt}, P.~R.~M., {Evans}, T.,
  {Fajardo-Acosta}, S., {Fowler}, J., {Gelino}, C., {Grillmair}, C., {Harbut},
  M., {Hoffman}, D., {Jarrett}, T., {Kirkpatrick}, J.~D., {Leisawitz}, D.,
  {Liu}, W., {Mainzer}, A., {Marsh}, K., {Masci}, F., {McCallon}, H.,
  {Padgett}, D., {Ressler}, M.~E., {Royer}, D., {Skrutskie}, M.~F., {Stanford},
  S.~A., {Wyatt}, P.~L., {Tholen}, D., {Tsai}, C.~W., {Wachter}, S.,
  {Wheelock}, S.~L., {Yan}, L., {Alles}, R., {Beck}, R., {Grav}, T., {Masiero},
  J., {McCollum}, B., {McGehee}, P., {Papin}, M., \& {Wittman}, M. 2012, 1

\bibitem[{{Farinella} {et~al.}(1998){Farinella}, {Vokrouhlicky}, \&
  {Hartmann}}]{Farinella.1998a}
{Farinella}, P., {Vokrouhlicky}, D., \& {Hartmann}, W.~K. 1998, \icarus, 132,
  378

\bibitem[{{Fowler} \& {Chillemi}(1992)}]{Fowler.1992a}
{Fowler}, J.~W. \& {Chillemi}, J.~R. 1992, The IRAS Minor Planet Survey,
  Phillips Laboratory, Hanscom AF Base, MA, Tech. Rep. PL-TR-92-2049, 17

\bibitem[{{Grav} {et~al.}(2012{\natexlab{a}}){Grav}, {Mainzer}, {Bauer},
  {Masiero}, {Spahr}, {McMillan}, {Walker}, {Cutri}, {Wright}, {Eisenhardt},
  {Blauvelt}, {DeBaun}, {Elsbury}, {Gautier}, {Gomillion}, {Hand}, \&
  {Wilkins}}]{Grav.2012a}
{Grav}, T., {Mainzer}, A.~K., {Bauer}, J., {Masiero}, J., {Spahr}, T.,
  {McMillan}, R.~S., {Walker}, R., {Cutri}, R., {Wright}, E., {Eisenhardt},
  P.~R., {Blauvelt}, E., {DeBaun}, E., {Elsbury}, D., {Gautier}, T.,
  {Gomillion}, S., {Hand}, E., \& {Wilkins}, A. 2012{\natexlab{a}}, \apj, 744,
  197

\bibitem[{{Grav} {et~al.}(2011){Grav}, {Mainzer}, {Bauer}, {Masiero}, {Spahr},
  {McMillan}, {Walker}, {Cutri}, {Wright}, {Eisenhardt}, {Blauvelt}, {DeBaun},
  {Elsbury}, {Gautier}, {Gomillion}, {Hand}, \& {Wilkins}}]{Grav.2011b}
{Grav}, T., {Mainzer}, A.~K., {Bauer}, J., {Masiero}, J., {Spahr}, T.,
  {McMillan}, R.~S., {Walker}, R., {Cutri}, R., {Wright}, E., {Eisenhardt},
  P.~R.~M., {Blauvelt}, E., {DeBaun}, E., {Elsbury}, D., {Gautier}, IV, T.,
  {Gomillion}, S., {Hand}, E., \& {Wilkins}, A. 2011, \apj, 742, 40

\bibitem[{{Grav} {et~al.}(2012{\natexlab{b}}){Grav}, {Mainzer}, {Bauer},
  {Masiero}, \& {Nugent}}]{Grav.2012b}
{Grav}, T., {Mainzer}, A.~K., {Bauer}, J.~M., {Masiero}, J.~R., \& {Nugent},
  C.~R. 2012{\natexlab{b}}, \apj, 759, 49

\bibitem[{{Harris}(2008)}]{Harris.2008a}
{Harris}, A. 2008, \nat, 453, 1178

\bibitem[{{Harris}(1998)}]{Harris.1998a}
{Harris}, A.~W. 1998, \icarus, 131, 291

\bibitem[{{Harris} \& {Young}(1989)}]{Harris.1989b}
{Harris}, A.~W. \& {Young}, J.~W. 1989, \icarus, 81, 314

\bibitem[{{Harris} {et~al.}(1989){Harris}, {Young}, {Contreiras}, {Dockweiler},
  {Belkora}, {Salo}, {Harris}, {Bowell}, {Poutanen}, {Binzel}, {Tholen}, \&
  {Wang}}]{Harris.1989a}
{Harris}, A.~W., {Young}, J.~W., {Contreiras}, L., {Dockweiler}, T., {Belkora},
  L., {Salo}, H., {Harris}, W.~D., {Bowell}, E., {Poutanen}, M., {Binzel},
  R.~P., {Tholen}, D.~J., \& {Wang}, S. 1989, \icarus, 81, 365

\bibitem[{{Hergenrother} \& {Whiteley}(2011)}]{Hergenrother.2011a}
{Hergenrother}, C.~W. \& {Whiteley}, R.~J. 2011, \icarus, 214, 194

\bibitem[{{Jedicke} \& {Metcalfe}(1998)}]{Jedicke.1998a}
{Jedicke}, R. \& {Metcalfe}, T.~S. 1998, \icarus, 131, 245

\bibitem[{{Lagerkvist} \& {Magnusson}(1990)}]{Lagerkvist.1990a}
{Lagerkvist}, C.-I. \& {Magnusson}, P. 1990, \aaps, 86, 119

\bibitem[{{Lebofsky} \& {Spencer}(1989)}]{Lebofsky.1989a}
{Lebofsky}, L.~A. \& {Spencer}, J.~R. 1989, in Asteroids II, ed. R.~P.
  {Binzel}, T.~{Gehrels}, \& M.~S. {Matthews}, 128--147

\bibitem[{{Lebofsky} {et~al.}(1978){Lebofsky}, {Veeder}, {Lebofsky}, \&
  {Matson}}]{Lebofsky.1978a}
{Lebofsky}, L.~A., {Veeder}, G.~J., {Lebofsky}, M.~J., \& {Matson}, D.~L. 1978,
  \icarus, 35, 336

\bibitem[{{Mainzer} {et~al.}(2011{\natexlab{a}}){Mainzer}, {Bauer}, {Grav},
  {Masiero}, {Cutri}, {Dailey}, {Eisenhardt}, {McMillan}, {Wright}, {Walker},
  {Jedicke}, {Spahr}, {Tholen}, {Alles}, {Beck}, {Brandenburg}, {Conrow},
  {Evans}, {Fowler}, {Jarrett}, {Marsh}, {Masci}, {McCallon}, {Wheelock},
  {Wittman}, {Wyatt}, {DeBaun}, {Elliott}, {Elsbury}, {Gautier}, {Gomillion},
  {Leisawitz}, {Maleszewski}, {Micheli}, \& {Wilkins}}]{Mainzer.2011a}
{Mainzer}, A., {Bauer}, J., {Grav}, T., {Masiero}, J., {Cutri}, R.~M.,
  {Dailey}, J., {Eisenhardt}, P., {McMillan}, R.~S., {Wright}, E., {Walker},
  R., {Jedicke}, R., {Spahr}, T., {Tholen}, D., {Alles}, R., {Beck}, R.,
  {Brandenburg}, H., {Conrow}, T., {Evans}, T., {Fowler}, J., {Jarrett}, T.,
  {Marsh}, K., {Masci}, F., {McCallon}, H., {Wheelock}, S., {Wittman}, M.,
  {Wyatt}, P., {DeBaun}, E., {Elliott}, G., {Elsbury}, D., {Gautier}, IV, T.,
  {Gomillion}, S., {Leisawitz}, D., {Maleszewski}, C., {Micheli}, M., \&
  {Wilkins}, A. 2011{\natexlab{a}}, \apj, 731, 53

\bibitem[{{Mainzer} {et~al.}(2011{\natexlab{b}}){Mainzer}, {Grav}, {Bauer},
  {Masiero}, {McMillan}, {Cutri}, {Walker}, {Wright}, {Eisenhardt}, {Tholen},
  {Spahr}, {Jedicke}, {Denneau}, {DeBaun}, {Elsbury}, {Gautier}, {Gomillion},
  {Hand}, {Mo}, {Watkins}, {Wilkins}, {Bryngelson}, {Del Pino Molina}, {Desai},
  {G{\'o}mez Camus}, {Hidalgo}, {Konstantopoulos}, {Larsen}, {Maleszewski},
  {Malkan}, {Mauduit}, {Mullan}, {Olszewski}, {Pforr}, {Saro}, {Scotti}, \&
  {Wasserman}}]{Mainzer.2011b}
{Mainzer}, A., {Grav}, T., {Bauer}, J., {Masiero}, J., {McMillan}, R.~S.,
  {Cutri}, R.~M., {Walker}, R., {Wright}, E., {Eisenhardt}, P., {Tholen},
  D.~J., {Spahr}, T., {Jedicke}, R., {Denneau}, L., {DeBaun}, E., {Elsbury},
  D., {Gautier}, T., {Gomillion}, S., {Hand}, E., {Mo}, W., {Watkins}, J.,
  {Wilkins}, A., {Bryngelson}, G.~L., {Del Pino Molina}, A., {Desai}, S.,
  {G{\'o}mez Camus}, M., {Hidalgo}, S.~L., {Konstantopoulos}, I., {Larsen},
  J.~A., {Maleszewski}, C., {Malkan}, M.~A., {Mauduit}, J.-C., {Mullan}, B.~L.,
  {Olszewski}, E.~W., {Pforr}, J., {Saro}, A., {Scotti}, J.~V., \& {Wasserman},
  L.~H. 2011{\natexlab{b}}, \apj, 743, 156

\bibitem[{{Mainzer} {et~al.}(2012{\natexlab{a}}){Mainzer}, {Grav}, {Masiero},
  {Bauer}, {Cutri}, {McMillan}, {Nugent}, {Tholen}, {Walker}, \&
  {Wright}}]{Mainzer.2012c}
{Mainzer}, A., {Grav}, T., {Masiero}, J., {Bauer}, J., {Cutri}, R.~M.,
  {McMillan}, R.~S., {Nugent}, C.~R., {Tholen}, D., {Walker}, R., \& {Wright},
  E.~L. 2012{\natexlab{a}}, \apjl, 760, L12

\bibitem[{{Mainzer} {et~al.}(2012{\natexlab{b}}){Mainzer}, {Grav}, {Masiero},
  {Bauer}, {McMillan}, {Giorgini}, {Spahr}, {Cutri}, {Tholen}, {Jedicke},
  {Walker}, {Wright}, \& {Nugent}}]{Mainzer.2012b}
{Mainzer}, A., {Grav}, T., {Masiero}, J., {Bauer}, J., {McMillan}, R.~S.,
  {Giorgini}, J., {Spahr}, T., {Cutri}, R.~M., {Tholen}, D.~J., {Jedicke}, R.,
  {Walker}, R., {Wright}, E., \& {Nugent}, C.~R. 2012{\natexlab{b}}, \apj, 752,
  110

\bibitem[{{Mainzer} {et~al.}(2011{\natexlab{c}}){Mainzer}, {Grav}, {Masiero},
  {Bauer}, {Wright}, {Cutri}, {McMillan}, {Cohen}, {Ressler}, \&
  {Eisenhardt}}]{Mainzer.2011c}
{Mainzer}, A., {Grav}, T., {Masiero}, J., {Bauer}, J., {Wright}, E., {Cutri},
  R.~M., {McMillan}, R.~S., {Cohen}, M., {Ressler}, M., \& {Eisenhardt}, P.
  2011{\natexlab{c}}, \apj, 736, 100

\bibitem[{{Mainzer} {et~al.}(2011{\natexlab{d}}){Mainzer}, {Grav}, {Masiero},
  {Bauer}, {Wright}, {Cutri}, {Walker}, \& {McMillan}}]{Mainzer.2011d}
{Mainzer}, A., {Grav}, T., {Masiero}, J., {Bauer}, J., {Wright}, E., {Cutri},
  R.~M., {Walker}, R., \& {McMillan}, R.~S. 2011{\natexlab{d}}, \apjl, 737, L9

\bibitem[{{Mainzer} {et~al.}(2012{\natexlab{c}}){Mainzer}, {Masiero}, {Grav},
  {Bauer}, {Tholen}, {McMillan}, {Wright}, {Spahr}, {Cutri}, {Walker}, {Mo},
  {Watkins}, {Hand}, \& {Maleszewski}}]{Mainzer.2012a}
{Mainzer}, A., {Masiero}, J., {Grav}, T., {Bauer}, J., {Tholen}, D.~J.,
  {McMillan}, R.~S., {Wright}, E., {Spahr}, T., {Cutri}, R.~M., {Walker}, R.,
  {Mo}, W., {Watkins}, J., {Hand}, E., \& {Maleszewski}, C. 2012{\natexlab{c}},
  \apj, 745, 7

\bibitem[{{Masci}(2013)}]{Masci.2013a}
{Masci}, F. 2013, {ICORE: Image Co-addition with Optional Resolution
  Enhancement}, astrophysics Source Code Library

\bibitem[{{Masci} \& {Fowler}(2009)}]{Masci.2009a}
{Masci}, F.~J. \& {Fowler}, J.~W. 2009, in Astronomical Society of the Pacific
  Conference Series, Vol. 411, Astronomical Data Analysis Software and Systems
  XVIII, ed. D.~A. {Bohlender}, D.~{Durand}, \& P.~{Dowler}, 67

\bibitem[{{Oberst} {et~al.}(2012){Oberst}, {Christou}, {Suggs}, {Moser},
  {Daubar}, {McEwen}, {Burchell}, {Kawamura}, {Hiesinger}, {W{\"u}nnemann},
  {Wagner}, \& {Robinson}}]{Oberst.2012a}
{Oberst}, J., {Christou}, A., {Suggs}, R., {Moser}, D., {Daubar}, I.~J.,
  {McEwen}, A.~S., {Burchell}, M., {Kawamura}, T., {Hiesinger}, H.,
  {W{\"u}nnemann}, K., {Wagner}, R., \& {Robinson}, M.~S. 2012, \planss, 74,
  179

\bibitem[{{Oszkiewicz} {et~al.}(2012){Oszkiewicz}, {Bowell}, {Wasserman},
  {Muinonen}, {Penttil{\"a}}, {Pieniluoma}, {Trilling}, \&
  {Thomas}}]{Oszkiewicz.2012a}
{Oszkiewicz}, D.~A., {Bowell}, E., {Wasserman}, L.~H., {Muinonen}, K.,
  {Penttil{\"a}}, A., {Pieniluoma}, T., {Trilling}, D.~E., \& {Thomas}, C.~A.
  2012, \icarus, 219, 283

\bibitem[{{Pravec} {et~al.}(2012){Pravec}, {Harris}, {Ku{\v s}nir{\'a}k},
  {Gal{\'a}d}, \& {Hornoch}}]{Pravec.2012a}
{Pravec}, P., {Harris}, A.~W., {Ku{\v s}nir{\'a}k}, P., {Gal{\'a}d}, A., \&
  {Hornoch}, K. 2012, \icarus, 221, 365

\bibitem[{{Pravec} {et~al.}(2008){Pravec}, {Harris}, {Vokrouhlick{\'y}},
  {Warner}, {Ku{\v s}nir{\'a}k}, {Hornoch}, {Pray}, {Higgins}, {Oey},
  {Gal{\'a}d}, {Gajdo{\v s}}, {Korno{\v s}}, {Vil{\'a}gi}, {Hus{\'a}rik},
  {Krugly}, {Shevchenko}, {Chiorny}, {Gaftonyuk}, {Cooney}, {Gross}, {Terrell},
  {Stephens}, {Dyvig}, {Reddy}, {Ries}, {Colas}, {Lecacheux}, {Durkee}, {Masi},
  {Koff}, \& {Goncalves}}]{Pravec.2008a}
{Pravec}, P., {Harris}, A.~W., {Vokrouhlick{\'y}}, D., {Warner}, B.~D., {Ku{\v
  s}nir{\'a}k}, P., {Hornoch}, K., {Pray}, D.~P., {Higgins}, D., {Oey}, J.,
  {Gal{\'a}d}, A., {Gajdo{\v s}}, {\v S}., {Korno{\v s}}, L., {Vil{\'a}gi}, J.,
  {Hus{\'a}rik}, M., {Krugly}, Y.~N., {Shevchenko}, V., {Chiorny}, V.,
  {Gaftonyuk}, N., {Cooney}, W.~R., {Gross}, J., {Terrell}, D., {Stephens},
  R.~D., {Dyvig}, R., {Reddy}, V., {Ries}, J.~G., {Colas}, F., {Lecacheux}, J.,
  {Durkee}, R., {Masi}, G., {Koff}, R.~A., \& {Goncalves}, R. 2008, \icarus,
  197, 497

\bibitem[{{Rossi} {et~al.}(2009){Rossi}, {Marzari}, \&
  {Scheeres}}]{Rossi.2009a}
{Rossi}, A., {Marzari}, F., \& {Scheeres}, D.~J. 2009, \icarus, 202, 95

\bibitem[{{Scheeres} {et~al.}(2010){Scheeres}, {Hartzell}, {S{\'a}nchez}, \&
  {Swift}}]{Scheeres.2010a}
{Scheeres}, D.~J., {Hartzell}, C.~M., {S{\'a}nchez}, P., \& {Swift}, M. 2010,
  \icarus, 210, 968

\bibitem[{{Spahr}(1998)}]{Spahr.1998a}
{Spahr}, T.~B. 1998, PhD thesis, University of Florida

\bibitem[{{Statler} {et~al.}(2013){Statler}, {Cotto-Figueroa}, {Riethmiller},
  \& {Sweeney}}]{Statler.2013a}
{Statler}, T.~S., {Cotto-Figueroa}, D., {Riethmiller}, D.~A., \& {Sweeney},
  K.~M. 2013, \icarus, 225, 141

\bibitem[{{Stevenson} {et~al.}(2013){Stevenson}, {Bauer}, {Kramer}, {Grav},
  {Mainzer}, \& {Masiero}}]{Stevenson.2013a}
{Stevenson}, R., {Bauer}, J., {Kramer}, E., {Grav}, T., {Mainzer}, A., \&
  {Masiero}, J. 2013, \apj, submitted

\bibitem[{{Stuart} \& {Binzel}(2004)}]{Stuart.2004a}
{Stuart}, J.~S. \& {Binzel}, R.~P. 2004, \icarus, 170, 295

\bibitem[{{Suggs} {et~al.}(2008){Suggs}, {Cooke}, {Suggs}, {Swift}, \&
  {Hollon}}]{Suggs.2008a}
{Suggs}, R.~M., {Cooke}, W.~J., {Suggs}, R.~J., {Swift}, W.~R., \& {Hollon}, N.
  2008, Earth Moon and Planets, 102, 293

\bibitem[{{Trilling} {et~al.}(2010){Trilling}, {Mueller}, {Hora}, {Harris},
  {Bhattacharya}, {Bottke}, {Chesley}, {Delbo}, {Emery}, {Fazio}, {Mainzer},
  {Penprase}, {Smith}, {Spahr}, {Stansberry}, \& {Thomas}}]{Trilling.2010a}
{Trilling}, D.~E., {Mueller}, M., {Hora}, J.~L., {Harris}, A.~W.,
  {Bhattacharya}, B., {Bottke}, W.~F., {Chesley}, S., {Delbo}, M., {Emery},
  J.~P., {Fazio}, G., {Mainzer}, A., {Penprase}, B., {Smith}, H.~A., {Spahr},
  T.~B., {Stansberry}, J.~A., \& {Thomas}, C.~A. 2010, \aj, 140, 770

\bibitem[{{Veeder} {et~al.}(1978){Veeder}, {Matson}, \& {Smith}}]{Veeder.1978a}
{Veeder}, G.~J., {Matson}, D.~L., \& {Smith}, J.~C. 1978, \aj, 83, 651

\bibitem[{{Warner} {et~al.}(2009){Warner}, {Harris}, \&
  {Pravec}}]{Warner.2009a}
{Warner}, B.~D., {Harris}, A.~W., \& {Pravec}, P. 2009, \icarus, 202, 134

\bibitem[{{Wright} {et~al.}(2010){Wright}, {Eisenhardt}, {Mainzer}, {Ressler},
  {Cutri}, {Jarrett}, {Kirkpatrick}, {Padgett}, {McMillan}, {Skrutskie},
  {Stanford}, {Cohen}, {Walker}, {Mather}, {Leisawitz}, {Gautier}, {McLean},
  {Benford}, {Lonsdale}, {Blain}, {Mendez}, {Irace}, {Duval}, {Liu}, {Royer},
  {Heinrichsen}, {Howard}, {Shannon}, {Kendall}, {Walsh}, {Larsen}, {Cardon},
  {Schick}, {Schwalm}, {Abid}, {Fabinsky}, {Naes}, \& {Tsai}}]{Wright.2010a}
{Wright}, E.~L., {Eisenhardt}, P.~R.~M., {Mainzer}, A.~K., {Ressler}, M.~E.,
  {Cutri}, R.~M., {Jarrett}, T., {Kirkpatrick}, J.~D., {Padgett}, D.,
  {McMillan}, R.~S., {Skrutskie}, M., {Stanford}, S.~A., {Cohen}, M., {Walker},
  R.~G., {Mather}, J.~C., {Leisawitz}, D., {Gautier}, III, T.~N., {McLean}, I.,
  {Benford}, D., {Lonsdale}, C.~J., {Blain}, A., {Mendez}, B., {Irace}, W.~R.,
  {Duval}, V., {Liu}, F., {Royer}, D., {Heinrichsen}, I., {Howard}, J.,
  {Shannon}, M., {Kendall}, M., {Walsh}, A.~L., {Larsen}, M., {Cardon}, J.~G.,
  {Schick}, S., {Schwalm}, M., {Abid}, M., {Fabinsky}, B., {Naes}, L., \&
  {Tsai}, C.-W. 2010, \aj, 140, 1868

\end{thebibliography}
\end{document}